\documentclass[preprint,showpacs,amsmath,amssymb]{revtex4-2}
\usepackage{amsmath,amssymb,latexsym,epsfig,graphics,epsf}
\usepackage{graphics}
\usepackage{float}
\usepackage{todonotes}

\begin{document}

\title{Predicting scaling properties from a single fluid configuration}
\date{\today}
\author{Thomas B. Schr{\o}der}
\email{tbs@ruc.dk}
\affiliation{Glass and Time, IMFUFA, Department of Science and Environment, Roskilde University, P.O. Box 260, DK-4000 Roskilde, Denmark}

\begin{abstract}
Time-dependent dynamical properties of a fluid can not be estimated from a single configuration without performing a simulation. Here we show, however, that the \emph{scaling properties} of both structure and dynamics can be predicted from a single configuration. The new method is demonstrated to work very well for equilibrium dynamics of the Kob-Andersen Binary Lennard-Jones mixture. Furthermore, the method is applied to isobaric cooling where the liquid falls out of equilibrium and forms a glass, demonstrating that the method requires neither equilibrium nor constant volume conditions to work, in contrast to existing methods.
\end{abstract}
\maketitle


How much can a single configuration tell us about the fluid under the conditions from which it was taken? To be specific, consider the positions of $N$ particles in 3 dimensions stored in a $3N$-dimensional vector, $\mathbf R \equiv \left( \mathbf r_1, \mathbf r_2, ..., \mathbf r_N \right)$, where $\mathbf r_i$ is the position of the i'th particle. Obviously, measures of the structure, such as the radial distribution function, can be estimated from a single configuration,  $\mathbf R$, with a precision that depends on $N$. Assuming knowledge of the Hamiltonian, also thermodynamic properties such as potential energy and pressure can be estimated with a per-particle error proportional to $1/\sqrt{N}$. On the other hand, time-dependent dynamical properties such as the mean-square displacement and intermediate scattering function can \emph{not} be estimated from a single configuration. This letter demonstrates, however, that the \emph{scaling properties} of both structure and dynamics can be predicted from a single configuration.

Scaling relations play an important role in physics. Rosenfeld's excess entropy scaling\cite{ros77,mitt06,chopra2010,abramson2011,dyre18_excessS,bell20} states that transport coefficients of fluids depend only on the excess entropy, $S_{ex} \equiv S - S_{ideal}$ ($S_{ideal}$ being the entropy of the ideal gas at the same density and temperature). 
Another scaling principle is the so-called power-law density scaling, stating that relaxation time and viscosity depend on temperature, $T$, and density, $\rho$, only via the combination $\rho^\gamma/T$, where $\gamma$ is a material-dependent scaling exponent \cite{tolle2001,Rolandetal2005,cos2009,sch09,frag2011} (For a more general scaling principle, see \cite{alba02, alba06}). The scaling requires the use of so-called reduced units, where  the unit of energy is given by $e_0\equiv k_BT$, the unit of length is given by $l_0\equiv\rho^{-1/3}$, and the unit of time is given by $t_0\equiv \rho^{-1/3}\sqrt{m/k_BT}$, where $m$ is a characteristic mass of the system. The described scaling properties -- including that they do {not} always  work  -- are explained by the isomorph theory \cite{paperIV,Boehling2012,SchroederDyre14}, which we will return to below. 


How can the dynamics in reduced units be the same at two state points $(\rho_1, T_1)$ and $(\rho_2, T_2)$? The simplest explanation is, that it is the same partial differential equation governing the dynamics at the two state-points\cite{dyre18_excessS}. Restricting ourselves to classical dynamics, Newtons second law can be written, $\mathbf F(\mathbf R) = \mathbf m {d^2 \mathbf R}/{d t^2}$, where $\mathbf F(\mathbf R)$ is the $3N$ dimensional vector containing the forces on the particles, and $\mathbf m$ is a diagonal matrix containing the relevant masses. 
Denoting reduced quantities by a tilde, the reduced force is given by: $\tilde{ \mathbf F} = {\mathbf F}/({e_0/l_0}) = {\mathbf F}/({\rho^{1/3}k_BT})$, and 
Newtons second law  becomes: $\tilde{\mathbf F}(\mathbf R) = \tilde{\mathbf m} {d^2 \tilde{ \mathbf {R}}}/{d \tilde{t}^2}$.
Thus, {if} the reduced force  depends only on the reduced coordinates,
then it is the {same} partial differential equation governing the dynamics at the two state points, which will result in  the same trajectory in reduced units $\tilde{ \mathbf {R}}(\tilde{t})$\cite{dyre18_excessS}, and thus the same mean-square displacement and intermediate scattering function in reduced units, as well as the same structure.

The proposed method works as follows. Given a configuration $\mathbf R_1$ with density $\rho_1$ and temperature $T_1$, we perform an affine scaling to density $\rho_2$, so that $\rho_2^{1/3}\mathbf R_2 = \rho_1^{1/3}\mathbf R_1$, i.e., the two configurations are the same in reduced units, $\tilde {\mathbf R}_2 = \tilde {\mathbf R}_1 $. Our aim is now to chose the temperature $T_2$ so that the reduced forces of $\mathbf R_1$ and $\mathbf R_2$, denoted $\tilde {\mathbf F}_1$ and $\tilde {\mathbf F}_2$ respectively, are as similar as possible. To this end we define an error function:
\begin{equation}
  Y \equiv \frac{(\tilde {\mathbf F}_2 - \tilde {\mathbf F}_1)^2}{ \tilde {\mathbf F}_1^2 + \tilde {\mathbf F}_2^2}.\label{eq:Y}
 \end{equation}
Taking the derivative of  Eq.~(\ref{eq:Y}) with respect to $T_2$ (which enters via the reduced units in $\tilde {\mathbf F}_2$), we find after straight forward manipulations that the minimum is located at:
 \begin{eqnarray}
   T_2 = \left(\frac{\rho_1}{\rho_2} \right)^{1/3}\frac{|\mathbf F_2|}{|\mathbf F_1|}  T_1,  \label{eq:T2min}
 \end{eqnarray}
 corresponding to choosing $T_2$ so that $|\tilde {\mathbf F}_1| = |\tilde {\mathbf F}_2|$.
 The value of the error function at the minimum is:
  \begin{equation}\label{eq:Ymin}
   Y = 1- R_{FF}, ~~~R_{FF} \equiv \frac{ \mathbf F_2\cdot \mathbf F_1}{|\mathbf F_2||\mathbf F_1|}
   =   \frac{ \tilde {\mathbf F}_2\cdot \tilde {\mathbf F}_1}{|\tilde {\mathbf F}_2||\tilde {\mathbf F}_1|}. 
  \end{equation}
  where $R_{FF}$ is the Pearson correlation coefficient of the force components, giving the cosine of the angle between  ${\mathbf F}_1$ and $ {\mathbf F}_2$ (and thus between $\tilde {\mathbf F}_1$ and $ \tilde {\mathbf F}_2$). 

\begin{figure}
	\vspace{1cm}
	\includegraphics[width=.45\textwidth]{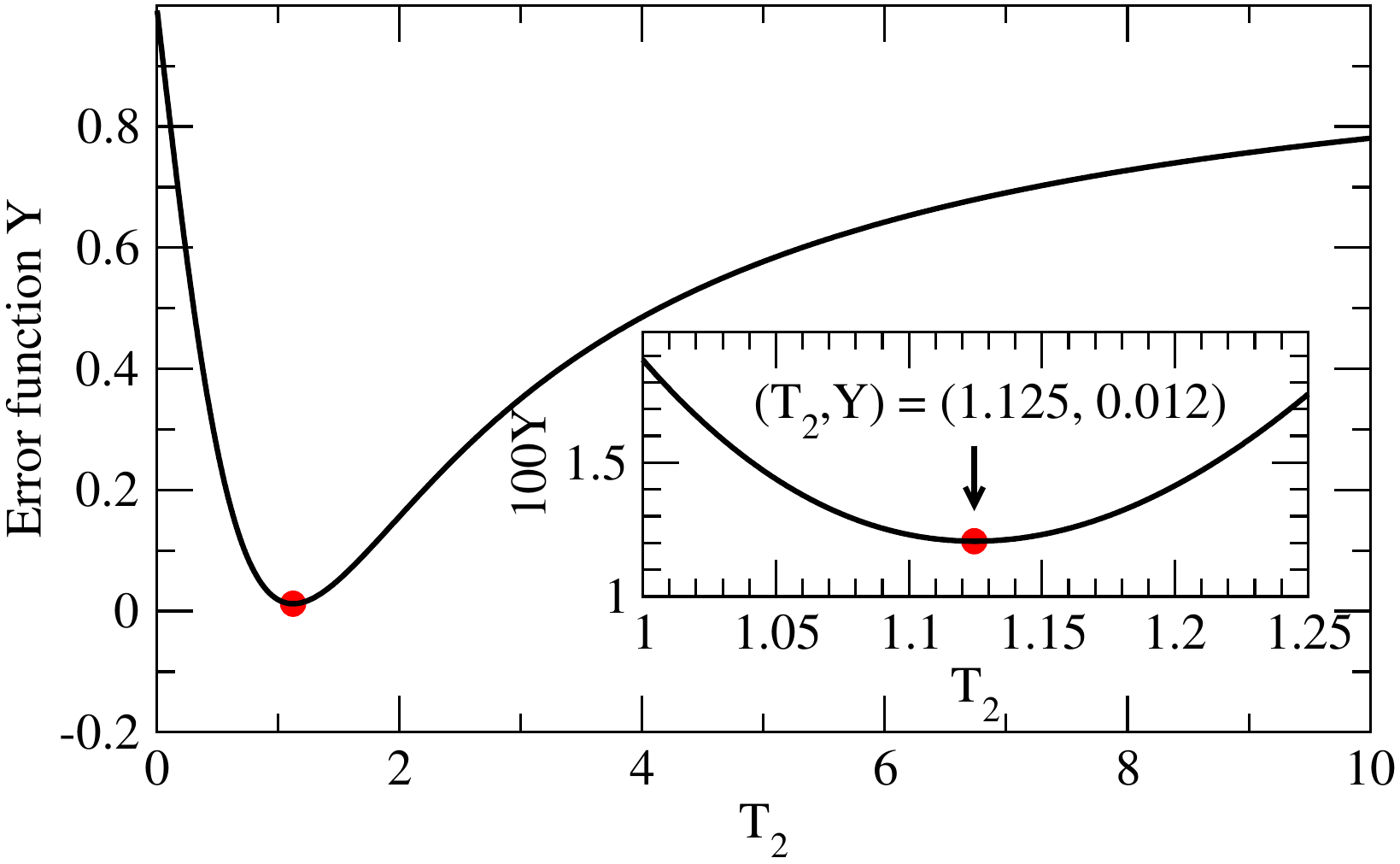}
	\caption{Error function, $Y$ (Eq.~(\ref{eq:Y})), for a single configuration of the Kob-Andersen binary LJ-mixture ($N=10000$). The configuration was taken from a NVT simulation at $(\rho_1,T_1) = (1.20, 0.450)$. Density was increased 20\%: $\rho_2 = 1.2\rho_1 = 1.44$. Inset: zoom on minimum, note scale.
		\label{fig:SingleConfErrorF}
	}
\end{figure}
 
In the following, the method is applied to the 80:20 Kob-Andersen binary Lennard-Jones mixture\cite{kob94}, a standard model in simulations of viscous liquids. NVT simulations using a Nose-Hoover thermostat with N=10000 particles were performed using RUMD\cite{RUMD}, an open source molecular dynamics package optimized for GPU-computing. 

Fig.~\ref{fig:SingleConfErrorF} shows the error function, Eq.~(\ref{eq:Y}), evaluated for a single configuration taken from an equilibrium simulation at $(\rho_1, T_1) = (1.20, 0.45)$. A 20\% increase in density was applied.
In experiments this would be considered a large density increase\cite{Ransom2019}.  
The minimum is given by $T_2 = 1.125$, and $Y=0.012$ corresponding to a Pearson correlation coefficient $R_{FF}=0.988$.

\begin{figure}
  	\includegraphics[width=.45\textwidth]{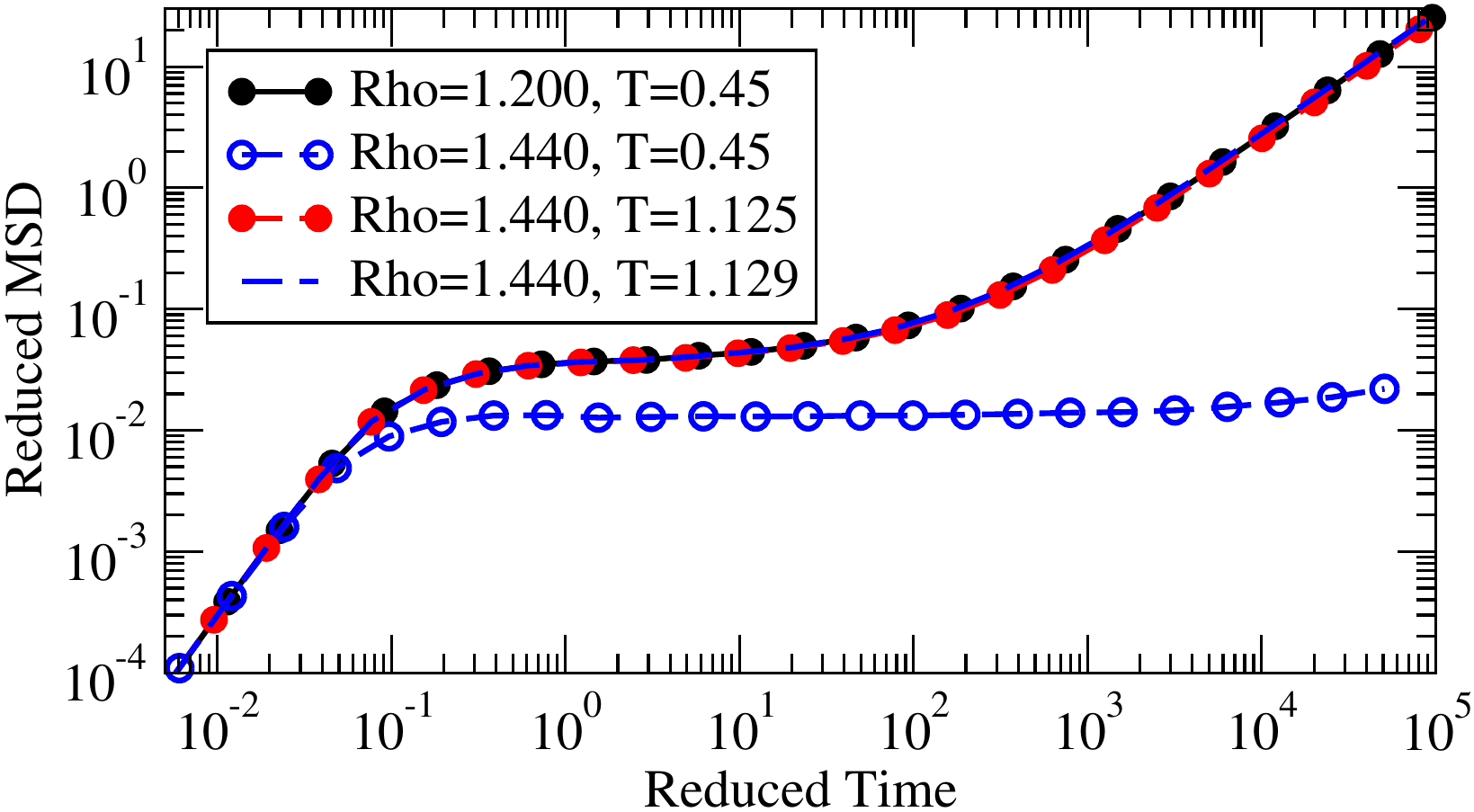}      
	\includegraphics[width=.45\textwidth]{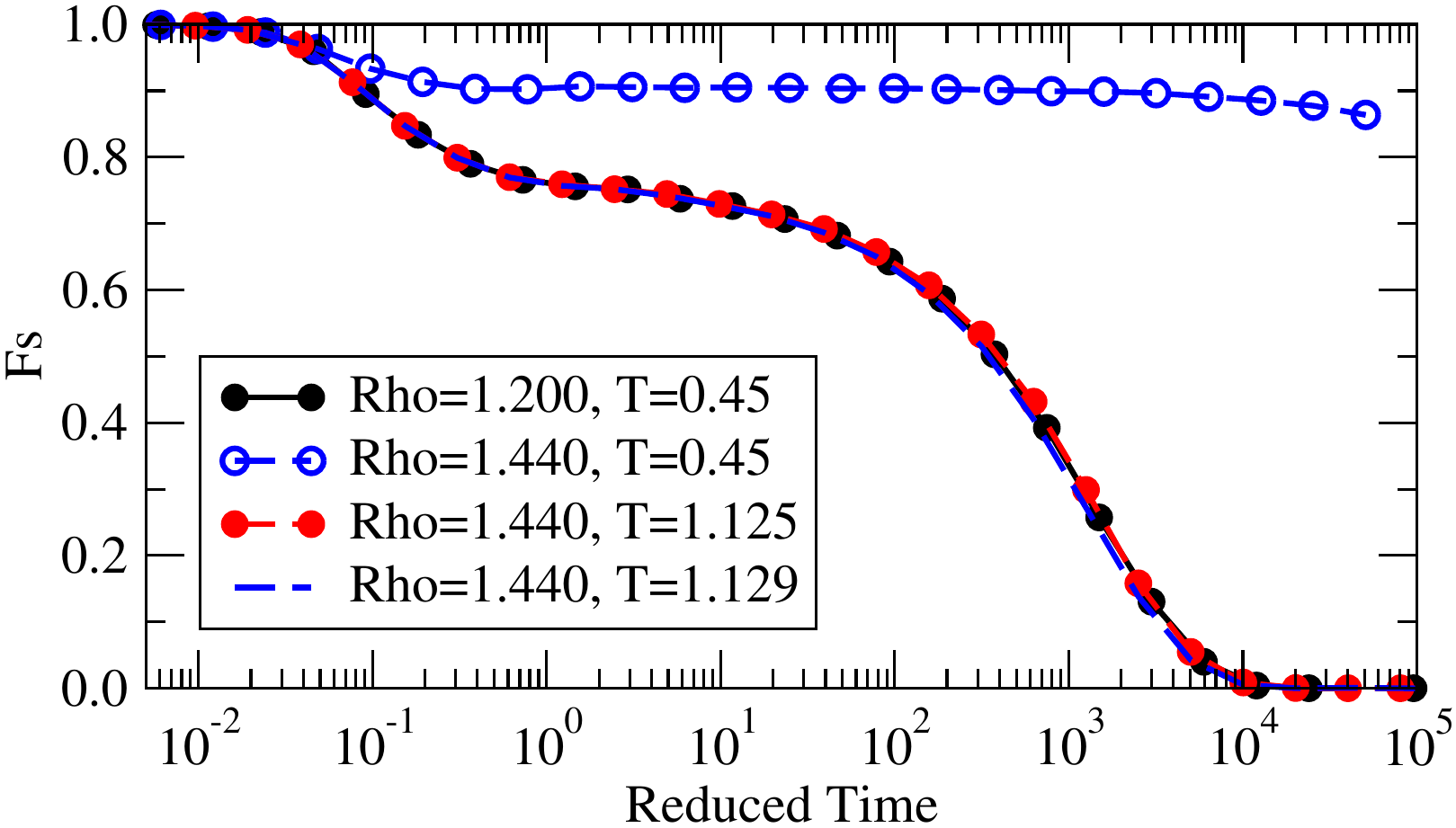}
        \caption{Testing invariance of dynamics predicted from a single configuration. Upper panel: mean-square displacement in reduced units. Lower panel: intermediate scattering function in reduced units ($q=7.25(\rho/1.2)^{1/3}$). In both cases, the dynamics at $(\rho_2,T_2) = (1.44, 1.125)$ is very close to that of the reference state point, $(\rho_1,T_1) = (1.20, 0.45)$. In contrast, the same density increase on the isotherm,  $(\rho,T) = (1.44, 0.45)$, drastically changes dynamics (open blue circles). Dynamics at $(\rho,T) = (1.44, 1.129)$ is shown as blue dashed lines. 
    	\label{fig:SingleConfDyn}
}
\end{figure}

The predicted invariance of the dynamics is tested in Fig.~\ref{fig:SingleConfDyn}. Results for the reference state point $(\rho_1,T_1) = (1.20, 0.45)$ (filled black circles) shows a plateau in the dynamics as characteristic for viscous liquids. The reduced dynamics at $(\rho_2,T_2) = (1.44, 1.125)$ (red filled circles) is to a very good approximation the same as at the reference state point. From this we conclude that the method works very well: from a single configuration we predicted a new state point at which the reduced dynamics is indistinguishable from that of the reference state point. The corresponding invariance of structure is shown in Fig. 1 in the supplemental material\cite{Supp}.

What if we had chosen a different configuration to apply the method to? Applying the method to 178 independent configurations gives a mean $T_2=1.1249$ with a standard deviation $0.0014$ (distribution shown in Fig. 2 in the supplemental material\cite{Supp}. Due to the strong temperature dependence of viscous liquids, we tested whether picking a configuration from the tail of this distribution, $T_2 = 1.129$,  alters the conclusion; the blue dashed lines in Fig.~\ref{fig:SingleConfDyn} shows that this is {not} the case. At more viscous reference state points, we expect the equilibrium dynamics to become even more temperature-dependent, and a larger sample size - or averaging over several configurations - would presumably be necessary. 

\begin{figure}
	\includegraphics[width=.23\textwidth]{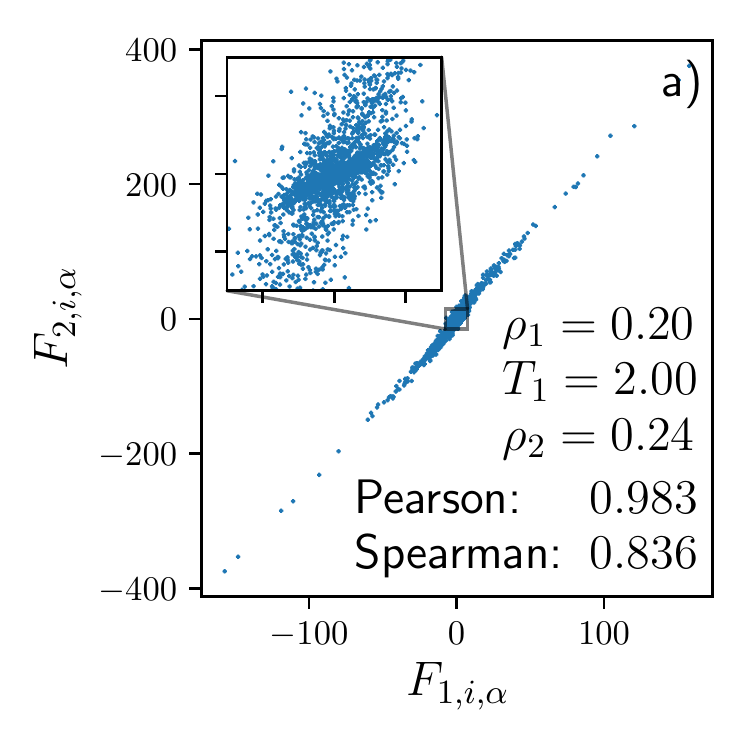}
	\includegraphics[width=.23\textwidth]{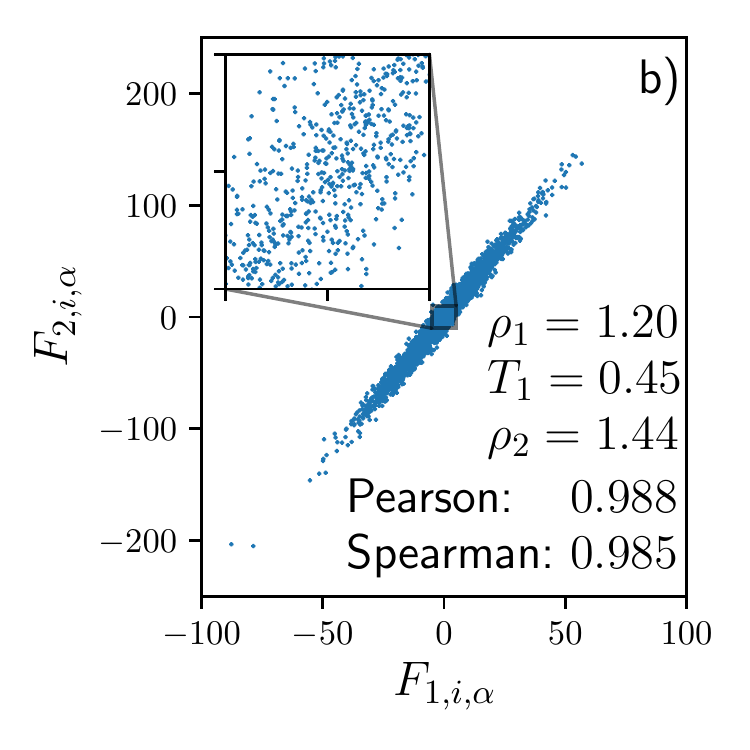}
	\includegraphics[width=.23\textwidth]{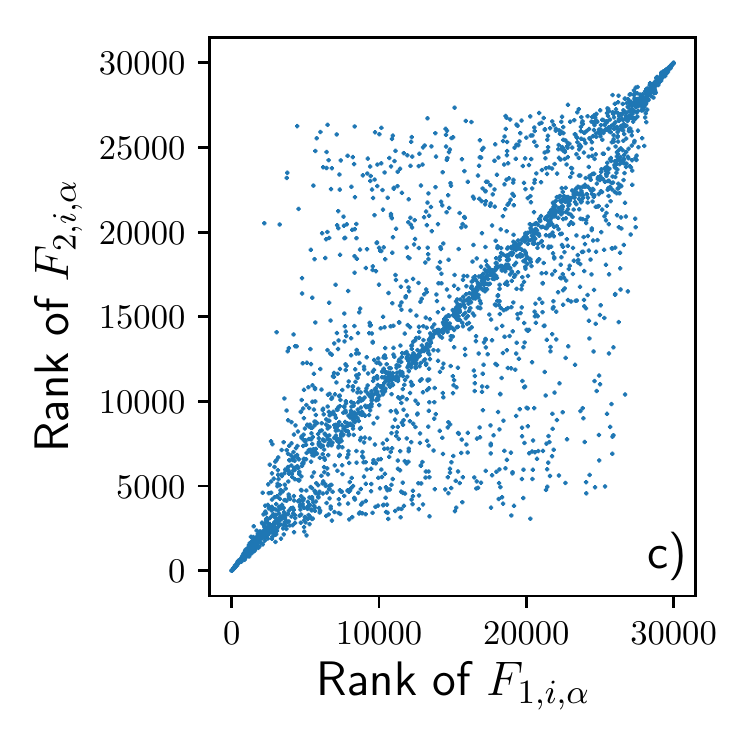}
	\includegraphics[width=.23\textwidth]{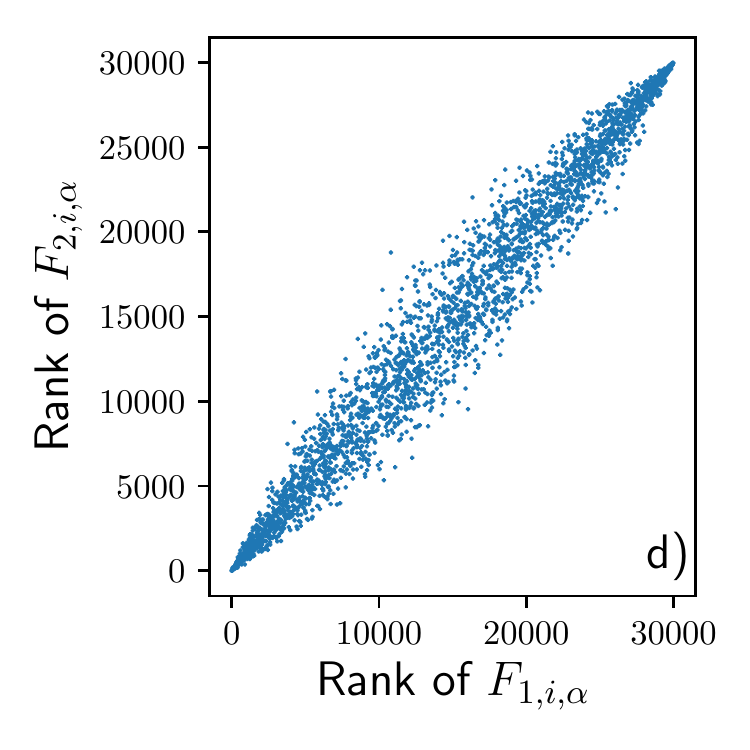}
	\caption{Force components, $F_{2,i,\alpha}$, of a scaled configuration versus force components, $F_{1,i,\alpha}$, of the same configuration before scaling. 10\% of data points displayed. a) Low density where scaling does not apply, $\rho_1=0.20, T_1=2.00, \rho_2=0.24$. b) High density where scaling does apply, $\rho_1=1.20, T_1=0.45, \rho_2=1.44$. c) and d) Same as above, except using the rank of the data.
		\label{fig:Fxyz}
	}
\end{figure}


Can the new method predict whether the scaling will work or not? A necessary requirement is that the force components before and after scaling are strongly correlated, i.e., that $ {\mathbf F}_1$ and ${\mathbf F}_2$ are close to being parallel, and thus  $\tilde {\mathbf F}_1 \approx \tilde {\mathbf F}_2$. 
Fig.~\ref{fig:Fxyz}a shows a scatter plot of the force components before and after scaling for a low density state point where scaling does not apply (see  Fig. 3 in the supplemental material\cite{Supp}). The inset reveals that a subset of small force components is correlated with a smaller slope, a feature that is not present (Fig.~\ref{fig:Fxyz}b) for the high density state point where scaling works (as demonstrated in Fig.~\ref{fig:SingleConfDyn}).
This difference does not lead to significantly different Pearson correlation coefficients ($0.983$ and $0.988$). In contrast, the Spearman correlation coefficient (the Pearson correlation coefficient of the rank of the data, plotted in Figs.~\ref{fig:Fxyz}c and d), is different: $0.836$ and $0.985$, respectively. More tests are needed, but a good criterion for expecting the scaling to work might be that the Pearson and Spearman correlation coefficients both should be larger than $0.95$. 


We conclude from the results presented above that when the scaling works, the two state point are indeed characterized by reduced forces to a good approximation depending only on the reduced coordinates. The simplest explanation for this is that the reduced potential surface is the same, except for an additative constant:

\begin{equation}
  \frac{U(\mathbf R_2)}{k_BT_2} = \frac{U(\mathbf R_1)}{k_BT_1} + C, ~~~\rho_2^{1/3}\mathbf R_2 = \rho_1^{1/3}\mathbf R_1 \label{Eq:isom_assumption}
\end{equation}

This is the basic assumption of the isomorph theory\cite{paperIV} (for a more general formulation, see Ref.\cite{SchroederDyre14}). 
For two state points fulfilling Eq.~(\ref{Eq:isom_assumption}), it is straightforward to show i) the reduced forces are the same, and therefore the dynamics is the same in reduced units, ii) the structure is the same in reduced units, and iii) the excess entropy, $S_{ex}$, is the same. 'Isomorphs' are curves in the phase diagram where any two points fulfill Eq.~(\ref{Eq:isom_assumption}) to a good approximation, and are thus characterized by approximate invariant reduced-unit structure and dynamics as well as excess entropy. 

In the context of the isomorph theory, the new force based method presented here is a new method for identifying isomorphs. For comparison, we will briefly describe the two by far most used existing methods, the $\gamma$-method, and the direct isomorph check. 


In the $\gamma$-method, a curve in the phase diagram with constant $S_{ex}$, i.e., a configurational adiabat, is traced out, using the general statistical mechanics identity\cite{paperIV}:
\begin{equation}
\gamma \equiv \left( \frac{\partial \ln T}{\partial \ln \rho} \right)_{S_{ex}} =
\frac{\left< \Delta W \Delta U\right>}{\left< \left( \Delta U\right)^2  \right>} \label{Eq:GammaFluct}
\end{equation}
where $\left< ... \right>$ denotes canonical (NVT) ensemble average, and $\Delta$ denotes deviation from the ensemble average, eg. $\Delta U \equiv U - \left< U \right>$. The right hand side of  this equation is evaluated from equilibrium NVT simulations, and the configurational adiabat is traced out by numerically solving the differential equation, Eq.~(\ref{Eq:GammaFluct}). 
The main disadvantage of this method is that it requires small steps in density, 
which can pose a practical problem in particular in viscous liquids where long simulations are needed to accurately evaluate the right hand side of Eq.~(\ref{Eq:GammaFluct}).

Eq.~(\ref{Eq:isom_assumption}) can be re-written as:
\begin{equation}
  U(\mathbf R_2) = \frac{T_2}{T_1} U(\mathbf R_1) + D, ~~~\rho_2^{1/3}\mathbf R_2 = \rho_1^{1/3}\mathbf R_1 \label{Eq:DIC}
\end{equation}
This is the basis of the direct isomorph check\cite{paperIV}: i) an equilibrium NVT simulation is performed at a state point $(\rho_1, T_1)$. ii) a number of configurations are scaled affinely to a new density $\rho_2$. iii) the potential energy of the scaled configurations, $U(\mathbf R_2)$, are plotted against the potential energy of the un-scaled configurations, $U(\mathbf R_1)$ in a scatter-plot. iv) for the isomorph theory to apply  $U(\mathbf R_2)$ and $U(\mathbf R_1)$ needs to be strongly correlating. v) the new temperature can be determined from the slope being equal to $T_2/T_1$.
The main advantage of this method is that large density changes can be performed from a single equilibrium NVT simulation. For Lennard-Jones type systems, the direct isomorph check can be performed analytically\cite{Ingebrigtsen2012b,Boehling2012}:
\begin{equation}
\frac{T_2}{T_1} = \left(\frac{\rho_2}{\rho_1}\right)^4\left(\frac{\gamma_1}{2} - 1 \right) 
-  \left(\frac{\rho_2}{\rho_1}\right)^2\left(\frac{\gamma_1}{2} - 2 \right), \label{Eq:DICanalytic}
\end{equation}
where $\gamma_1$ is Eq.~(\ref{Eq:GammaFluct}) evaluated at the reference state point.
\begin{figure}
	\includegraphics[width=.45\textwidth]{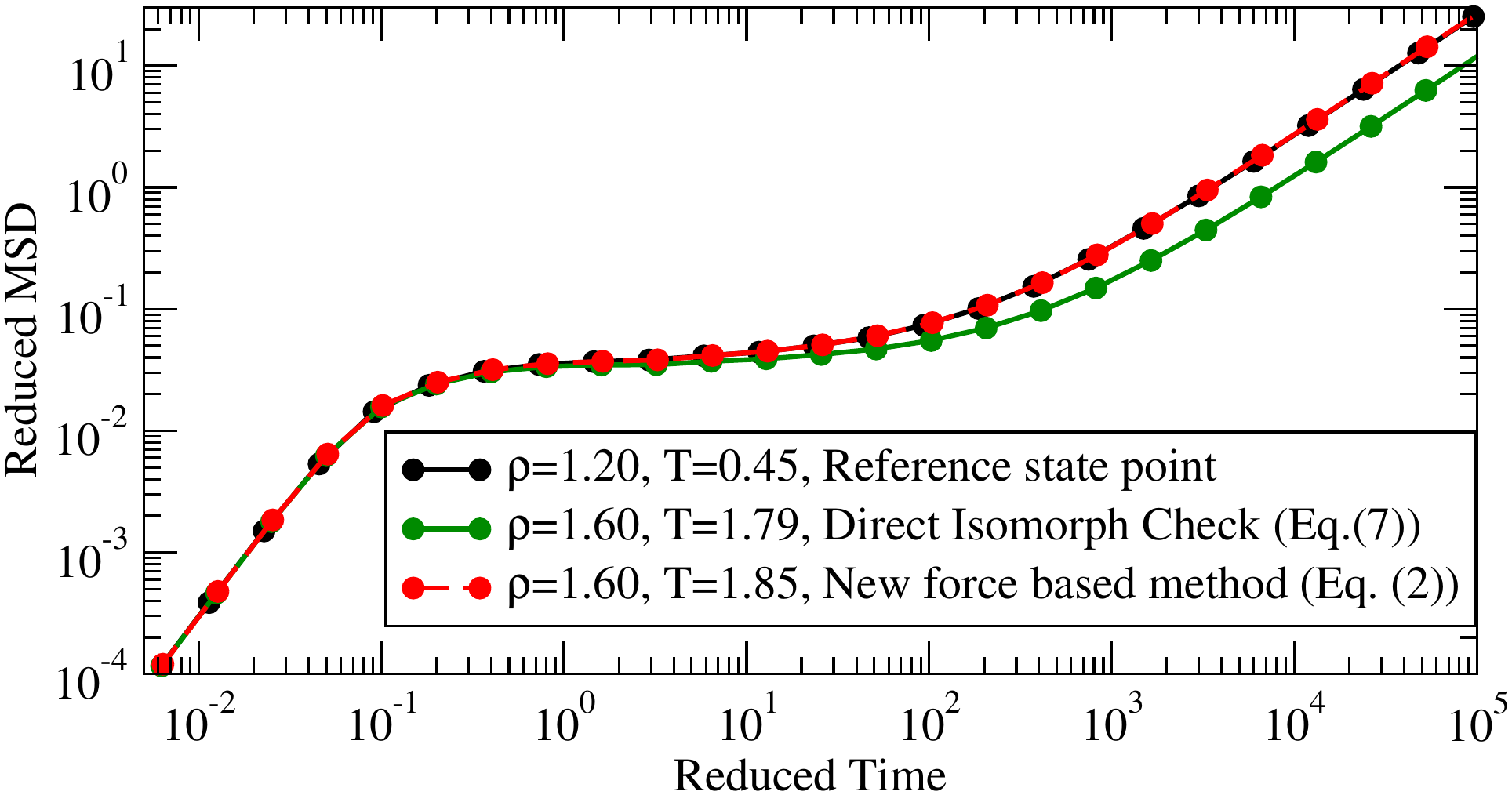}      
	\caption{Mean squared displacement in reduced units. Black circles: Reference state point (same as in Fig.~\ref{fig:SingleConfDyn}). Green circles: Direct isomorph check (Eq.~(\ref{Eq:DICanalytic})) using $\gamma_1 = 5.167$, resulting in dynamics approximately a factor two too slow. Red circles: The force method (Eq.~(\ref{eq:T2min})). 178 independent configurations give an average $T_2=1.851$ with standard deviation $0.003$. 
		\label{fig:SingleConfDyn1.6}
	}
\end{figure}


In Fig.~\ref{fig:SingleConfDyn1.6} the direct isomorph check (Eq.~(\ref{Eq:DICanalytic})) is compared to the new force based method (Eq.~(\ref{eq:T2min})), using $\rho_2=1.60$, i.e., a 33\% density increase. The new method clearly outperforms the direct isomorph check (which is known to work well for less viscous systems \cite{SchroederDyre14,Costigliola2018,Costigliola2019,Rahman2021}).


\begin{figure}
	\includegraphics[width=.45\textwidth]{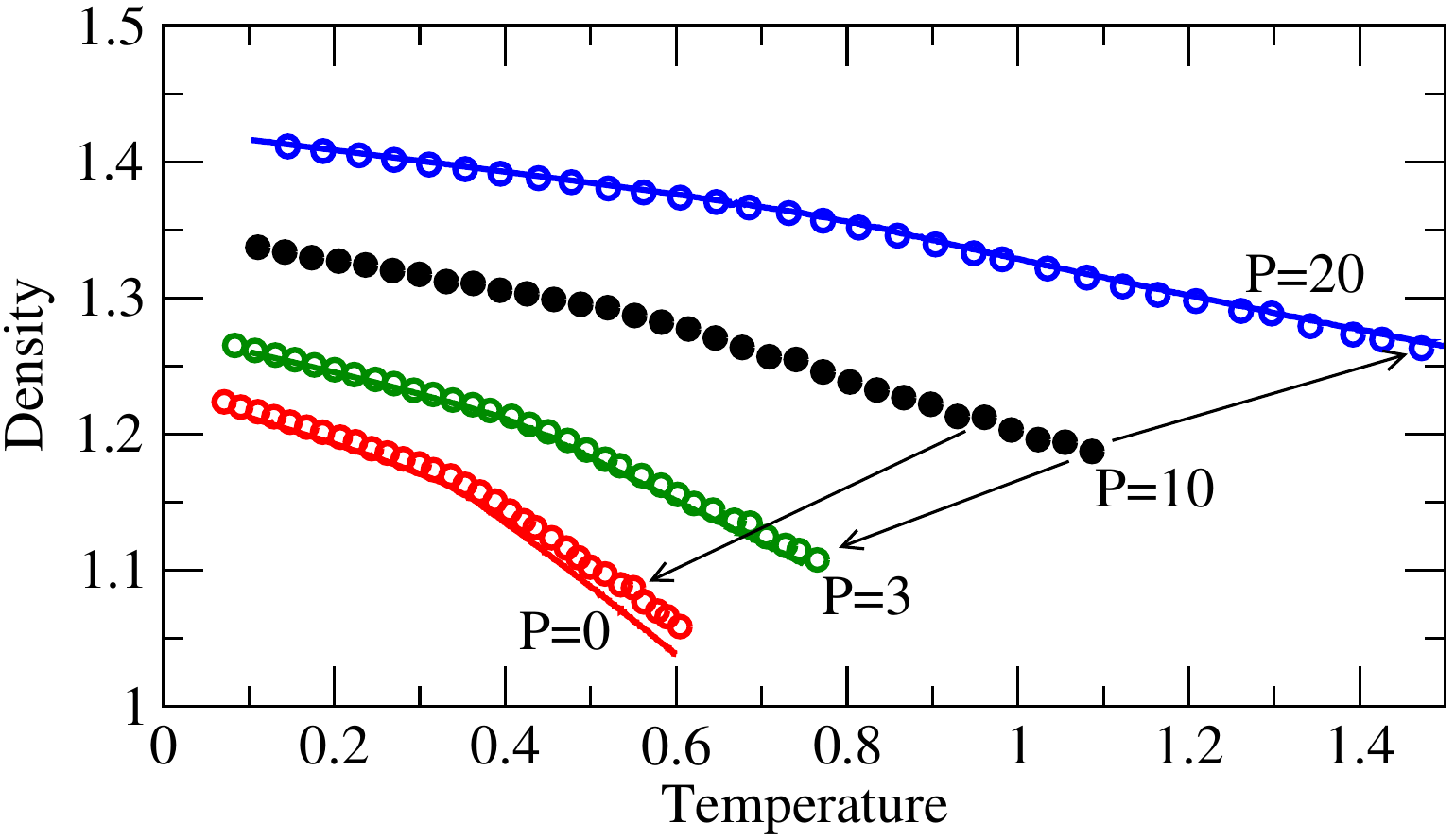} 
	\caption{Density, $\rho$, versus temperature, $T$, for isobaric cooling curves. Predictions for  cooling curves at $P=20,3$, and $0$ (blue, green, and red open circles) was calculated from the $P=10$ cooling curve (filled black circles), as described in the text. Good agreement is found with actual simulated cooling curves (blue, green, and red lines).
		\label{fig:SingleConfCooling}
	}
\end{figure}

The new method can be applied to a single configuration, which does {not} have to be from a constant volume simulation, nor does it have to be in equilibrium. In the following we will show-case an application illustrating the last two points - to our knowledge no other method has these possibilities.

When a glass-forming liquid is cooled continuously it eventually falls out of equilibrium and forms a glass, i.e., a glass transition is observed at the glass temperature, $T_g$. 
In Fig.~{\ref{fig:SingleConfCooling}} the black circles shows the density as a function of temperature during continuous cooling ($dT/dt = -2.0\cdot 10^{-6}$) at pressure $P=10$ (MD-units). To be specific, it is the density of individual configurations visited during the cooling that is plotted. This leads to some scatter in the data, but a glass transition around $T_g \approx 0.6$ is still clearly observed from the change of slope.

From the $P=10$ cooling curve, the 'isomorphic' cooling curve for $P=20$ is predicted as follows: Each of the stored $P=10$ configurations (black circles in Fig.~\ref{fig:SingleConfCooling}) are scaled to a higher density, the 'isomorphic' temperature, $T_2$, is evaluated from Eq.~(\ref{eq:T2min}), and the pressure is evaluated from $P = \rho_2 k_BT_2 + \rho_2 W_2/N$, where $W_2$ is the virial of the scaled configuration. This procedure is repeated, adjusting the density until 
the desired pressure $P=20$ is achieved with satisfactory precision. The resulting $(T,\rho)$ pairs are plotted as blue circles in   Fig.~\ref{fig:SingleConfCooling}. The 'isomorphic' cooling rate is estimated by plotting the predicted temperatures versus the time associated with the corresponding P=10  configurations, and fitting a straight line in the region corresponding to the glass transition (see  Fig. 4 in the supplemental material\cite{Supp}). The resulting cooling rate, $dT/dt = -2.71\cdot 10^{-6}$ was used for simulating a P=20 cooling curve shown as the full blue line in Fig.~\ref{fig:SingleConfCooling} (a running average over $10^3$ time units was applied).
Corresponding results for $P=3$ ($dT/dt = -1.40\cdot 10^{-6}$) and $P=0$ ($dT/dt = -1.11\cdot 10^{-6}$) are shown in green and red respectively. 
Deviations are seen at the lowest pressure $P=0$, but in general quite good agreement is observed between the actual simulated cooling curves (full lines) and the predicted cooling curves (open circles). This confirms that the force method can indeed be used without constant volume, and out of equilibrium, as argued above.

To summarize, we have presented an easily applicable method to predict scaling properties of fluids. The method was demonstrated to work very well, leading to the intriguing consequence that information about scaling properties is contained in individual configurations. This is in stark contrast, e.g., to Rosenfeld's excess entropy scaling\cite{ros77,mitt06,dyre18_excessS}, where the excess entropy usually is calculated by thermodynamic integration, requiring equilibrium data along paths in the phase diagram. A further feature of the new method presented here is that it does not require constant volume, nor equilibrium conditions, in contrast to existing methods.

	The author thanks Jeppe Dyre, Nicholas Bailey, and Lorenzo Costigliola for fruitful discussions. This work was supported by a research Grant (No. 00023189) from VILLUM FONDEN.

\bibliography{Single}

\begin{thebibliography}{24}%
\makeatletter
\providecommand \@ifxundefined [1]{%
 \@ifx{#1\undefined}
}%
\providecommand \@ifnum [1]{%
 \ifnum #1\expandafter \@firstoftwo
 \else \expandafter \@secondoftwo
 \fi
}%
\providecommand \@ifx [1]{%
 \ifx #1\expandafter \@firstoftwo
 \else \expandafter \@secondoftwo
 \fi
}%
\providecommand \natexlab [1]{#1}%
\providecommand \enquote  [1]{``#1''}%
\providecommand \bibnamefont  [1]{#1}%
\providecommand \bibfnamefont [1]{#1}%
\providecommand \citenamefont [1]{#1}%
\providecommand \href@noop [0]{\@secondoftwo}%
\providecommand \href [0]{\begingroup \@sanitize@url \@href}%
\providecommand \@href[1]{\@@startlink{#1}\@@href}%
\providecommand \@@href[1]{\endgroup#1\@@endlink}%
\providecommand \@sanitize@url [0]{\catcode `\\12\catcode `\$12\catcode
  `\&12\catcode `\#12\catcode `\^12\catcode `\_12\catcode `\%12\relax}%
\providecommand \@@startlink[1]{}%
\providecommand \@@endlink[0]{}%
\providecommand \url  [0]{\begingroup\@sanitize@url \@url }%
\providecommand \@url [1]{\endgroup\@href {#1}{\urlprefix }}%
\providecommand \urlprefix  [0]{URL }%
\providecommand \Eprint [0]{\href }%
\providecommand \doibase [0]{https://doi.org/}%
\providecommand \selectlanguage [0]{\@gobble}%
\providecommand \bibinfo  [0]{\@secondoftwo}%
\providecommand \bibfield  [0]{\@secondoftwo}%
\providecommand \translation [1]{[#1]}%
\providecommand \BibitemOpen [0]{}%
\providecommand \bibitemStop [0]{}%
\providecommand \bibitemNoStop [0]{.\EOS\space}%
\providecommand \EOS [0]{\spacefactor3000\relax}%
\providecommand \BibitemShut  [1]{\csname bibitem#1\endcsname}%
\let\auto@bib@innerbib\@empty
\bibitem [{\citenamefont {Rosenfeld}(1977)}]{ros77}%
  \BibitemOpen
  \bibfield  {author} {\bibinfo {author} {\bibfnamefont {Y.}~\bibnamefont
  {Rosenfeld}},\ }\bibfield  {title} {\bibinfo {title} {Relation between the
  transport coefficients and the internal entropy of simple systems},\ }\href
  {https://doi.org/10.1103/PhysRevA.15.2545} {\bibfield  {journal} {\bibinfo
  {journal} {Phys. Rev. A}\ }\textbf {\bibinfo {volume} {15}},\ \bibinfo
  {pages} {2545} (\bibinfo {year} {1977})}\BibitemShut {NoStop}%
\bibitem [{\citenamefont {Mittal}\ \emph {et~al.}(2006)\citenamefont {Mittal},
  \citenamefont {Errington},\ and\ \citenamefont {Truskett}}]{mitt06}%
  \BibitemOpen
  \bibfield  {author} {\bibinfo {author} {\bibfnamefont {J.}~\bibnamefont
  {Mittal}}, \bibinfo {author} {\bibfnamefont {J.~R.}\ \bibnamefont
  {Errington}},\ and\ \bibinfo {author} {\bibfnamefont {T.~M.}\ \bibnamefont
  {Truskett}},\ }\bibfield  {title} {\bibinfo {title} {Thermodynamics predicts
  how confinement modifies the dynamics of the equilibrium hard-sphere fluid},\
  }\href {https://doi.org/10.1103/PhysRevLett.96.177804} {\bibfield  {journal}
  {\bibinfo  {journal} {Phys. Rev. Lett.}\ }\textbf {\bibinfo {volume} {96}},\
  \bibinfo {pages} {177804} (\bibinfo {year} {2006})}\BibitemShut {NoStop}%
\bibitem [{\citenamefont {Chopra}\ \emph {et~al.}(2010)\citenamefont {Chopra},
  \citenamefont {Truskett},\ and\ \citenamefont {Errington}}]{chopra2010}%
  \BibitemOpen
  \bibfield  {author} {\bibinfo {author} {\bibfnamefont {R.}~\bibnamefont
  {Chopra}}, \bibinfo {author} {\bibfnamefont {T.~M.}\ \bibnamefont
  {Truskett}},\ and\ \bibinfo {author} {\bibfnamefont {J.~R.}\ \bibnamefont
  {Errington}},\ }\bibfield  {title} {\bibinfo {title} {Excess entropy scaling
  of dynamic quantities for fluids of dumbbell-shaped particles},\ }\href
  {https://doi.org/10.1063/1.3477767} {\bibfield  {journal} {\bibinfo
  {journal} {The Journal of Chemical Physics}\ }\textbf {\bibinfo {volume}
  {133}},\ \bibinfo {pages} {104506} (\bibinfo {year} {2010})}\BibitemShut
  {NoStop}%
\bibitem [{\citenamefont {Abramson}(2011)}]{abramson2011}%
  \BibitemOpen
  \bibfield  {author} {\bibinfo {author} {\bibfnamefont {E.~H.}\ \bibnamefont
  {Abramson}},\ }\bibfield  {title} {\bibinfo {title} {Viscosity of argon to
  5 gpa and 673 k},\ }\href {https://doi.org/10.1080/08957959.2011.625554}
  {\bibfield  {journal} {\bibinfo  {journal} {High Pressure Res.}\ }\textbf
  {\bibinfo {volume} {31}},\ \bibinfo {pages} {544} (\bibinfo {year}
  {2011})}\BibitemShut {NoStop}%
\bibitem [{\citenamefont {Dyre}(2018)}]{dyre18_excessS}%
  \BibitemOpen
  \bibfield  {author} {\bibinfo {author} {\bibfnamefont {J.~C.}\ \bibnamefont
  {Dyre}},\ }\bibfield  {title} {\bibinfo {title} {Perspective: Excess-entropy
  scaling},\ }\href {https://doi.org/10.1063/1.5055064} {\bibfield  {journal}
  {\bibinfo  {journal} {J. Chem. Phys.}\ }\textbf {\bibinfo {volume} {149}},\
  \bibinfo {pages} {210901} (\bibinfo {year} {2018})}\BibitemShut {NoStop}%
\bibitem [{\citenamefont {Bell}\ \emph {et~al.}(2020)\citenamefont {Bell},
  \citenamefont {Dyre},\ and\ \citenamefont {Ingebrigtsen}}]{bell20}%
  \BibitemOpen
  \bibfield  {author} {\bibinfo {author} {\bibfnamefont {I.~H.}\ \bibnamefont
  {Bell}}, \bibinfo {author} {\bibfnamefont {J.~C.}\ \bibnamefont {Dyre}},\
  and\ \bibinfo {author} {\bibfnamefont {T.~S.}\ \bibnamefont {Ingebrigtsen}},\
  }\bibfield  {title} {\bibinfo {title} {Excess-entropy scaling in supercooled
  binary mixtures},\ }\href {https://doi.org/10.1038/s41467-020-17948-1}
  {\bibfield  {journal} {\bibinfo  {journal} {Nat. Commun.}\ }\textbf {\bibinfo
  {volume} {11}},\ \bibinfo {pages} {4300} (\bibinfo {year}
  {2020})}\BibitemShut {NoStop}%
\bibitem [{\citenamefont {T\"olle}(2001)}]{tolle2001}%
  \BibitemOpen
  \bibfield  {author} {\bibinfo {author} {\bibfnamefont {A.}~\bibnamefont
  {T\"olle}},\ }\bibfield  {title} {\bibinfo {title} {Neutron scattering
  studies of the model glass former ortho-terphenyl},\ }\href
  {http://stacks.iop.org/0034-4885/64/i=11/a=203} {\bibfield  {journal}
  {\bibinfo  {journal} {Rep. Prog. Phys.}\ }\textbf {\bibinfo {volume} {64}},\
  \bibinfo {pages} {1473} (\bibinfo {year} {2001})}\BibitemShut {NoStop}%
\bibitem [{\citenamefont {Roland}\ \emph {et~al.}(2005)\citenamefont {Roland},
  \citenamefont {Hensel-Bielowka}, \citenamefont {Paluch},\ and\ \citenamefont
  {Casalini}}]{Rolandetal2005}%
  \BibitemOpen
  \bibfield  {author} {\bibinfo {author} {\bibfnamefont {C.~M.}\ \bibnamefont
  {Roland}}, \bibinfo {author} {\bibfnamefont {S.}~\bibnamefont
  {Hensel-Bielowka}}, \bibinfo {author} {\bibfnamefont {M.}~\bibnamefont
  {Paluch}},\ and\ \bibinfo {author} {\bibfnamefont {R.}~\bibnamefont
  {Casalini}},\ }\bibfield  {title} {\bibinfo {title} {Supercooled dynamics of
  glass-forming liquids and polymers under hydrostatic pressure},\ }\href
  {http://stacks.iop.org/0034-4885/68/i=6/a=R03} {\bibfield  {journal}
  {\bibinfo  {journal} {Rep. Prog. Phys.}\ }\textbf {\bibinfo {volume} {68}},\
  \bibinfo {pages} {1405} (\bibinfo {year} {2005})}\BibitemShut {NoStop}%
\bibitem [{\citenamefont {Coslovich}\ and\ \citenamefont
  {Roland}(2009)}]{cos2009}%
  \BibitemOpen
  \bibfield  {author} {\bibinfo {author} {\bibfnamefont {D.}~\bibnamefont
  {Coslovich}}\ and\ \bibinfo {author} {\bibfnamefont {C.~M.}\ \bibnamefont
  {Roland}},\ }\bibfield  {title} {\bibinfo {title} {Pressure-energy
  correlations and thermodynamic scaling in viscous lennard-jones liquids},\
  }\href {https://doi.org/10.1063/1.3054635} {\bibfield  {journal} {\bibinfo
  {journal} {J. Chem. Phys.}\ }\textbf {\bibinfo {volume} {130}},\ \bibinfo
  {pages} {014508} (\bibinfo {year} {2009})}\BibitemShut {NoStop}%
\bibitem [{\citenamefont {Schr\o{}der}\ \emph {et~al.}(2009)\citenamefont
  {Schr\o{}der}, \citenamefont {Pedersen}, \citenamefont {Bailey},
  \citenamefont {Toxvaerd},\ and\ \citenamefont {Dyre}}]{sch09}%
  \BibitemOpen
  \bibfield  {author} {\bibinfo {author} {\bibfnamefont {T.~B.}\ \bibnamefont
  {Schr\o{}der}}, \bibinfo {author} {\bibfnamefont {U.~R.}\ \bibnamefont
  {Pedersen}}, \bibinfo {author} {\bibfnamefont {N.~P.}\ \bibnamefont
  {Bailey}}, \bibinfo {author} {\bibfnamefont {S.}~\bibnamefont {Toxvaerd}},\
  and\ \bibinfo {author} {\bibfnamefont {J.~C.}\ \bibnamefont {Dyre}},\
  }\bibfield  {title} {\bibinfo {title} {Hidden scale invariance in molecular
  van der waals liquids: A simulation study},\ }\href
  {https://doi.org/10.1103/PhysRevE.80.041502} {\bibfield  {journal} {\bibinfo
  {journal} {Phys. Rev. E}\ }\textbf {\bibinfo {volume} {80}},\ \bibinfo
  {pages} {041502} (\bibinfo {year} {2009})}\BibitemShut {NoStop}%
\bibitem [{\citenamefont {Fragiadakis}\ and\ \citenamefont
  {Roland}(2011)}]{frag2011}%
  \BibitemOpen
  \bibfield  {author} {\bibinfo {author} {\bibfnamefont {D.}~\bibnamefont
  {Fragiadakis}}\ and\ \bibinfo {author} {\bibfnamefont {C.~M.}\ \bibnamefont
  {Roland}},\ }\bibfield  {title} {\bibinfo {title} {On the density scaling of
  liquid dynamics},\ }\href {https://doi.org/10.1063/1.3532545} {\bibfield
  {journal} {\bibinfo  {journal} {J. Chem. Phys.}\ }\textbf {\bibinfo {volume}
  {134}},\ \bibinfo {pages} {044504} (\bibinfo {year} {2011})}\BibitemShut
  {NoStop}%
\bibitem [{\citenamefont {Alba-Simionesco}\ \emph {et~al.}(2002)\citenamefont
  {Alba-Simionesco}, \citenamefont {Kivelson},\ and\ \citenamefont
  {Tarjus}}]{alba02}%
  \BibitemOpen
  \bibfield  {author} {\bibinfo {author} {\bibfnamefont {C.}~\bibnamefont
  {Alba-Simionesco}}, \bibinfo {author} {\bibfnamefont {D.}~\bibnamefont
  {Kivelson}},\ and\ \bibinfo {author} {\bibfnamefont {G.}~\bibnamefont
  {Tarjus}},\ }\bibfield  {title} {\bibinfo {title} {Temperature, density, and
  pressure dependence of relaxation times in supercooled liquids},\ }\href
  {https://doi.org/10.1063/1.1452724} {\bibfield  {journal} {\bibinfo
  {journal} {J. Chem. Phys.}\ }\textbf {\bibinfo {volume} {116}},\ \bibinfo
  {pages} {5033} (\bibinfo {year} {2002})}\BibitemShut {NoStop}%
\bibitem [{\citenamefont {Alba-Simionesco}\ and\ \citenamefont
  {Tarjus}(2006)}]{alba06}%
  \BibitemOpen
  \bibfield  {author} {\bibinfo {author} {\bibfnamefont {C.}~\bibnamefont
  {Alba-Simionesco}}\ and\ \bibinfo {author} {\bibfnamefont {G.}~\bibnamefont
  {Tarjus}},\ }\bibfield  {title} {\bibinfo {title} {Temperature versus density
  effects in glassforming liquids and polymers: A scaling hypothesis and its
  consequences},\ }\href
  {https://doi.org/https://doi.org/10.1016/j.jnoncrysol.2006.05.037} {\bibfield
   {journal} {\bibinfo  {journal} {J. Non-Cryst. Solids}\ }\textbf {\bibinfo
  {volume} {352}},\ \bibinfo {pages} {4888} (\bibinfo {year}
  {2006})}\BibitemShut {NoStop}%
\bibitem [{\citenamefont {Gnan}\ \emph {et~al.}(2009)\citenamefont {Gnan},
  \citenamefont {Schr\o{}der}, \citenamefont {Pedersen}, \citenamefont
  {Bailey},\ and\ \citenamefont {Dyre}}]{paperIV}%
  \BibitemOpen
  \bibfield  {author} {\bibinfo {author} {\bibfnamefont {N.}~\bibnamefont
  {Gnan}}, \bibinfo {author} {\bibfnamefont {T.~B.}\ \bibnamefont
  {Schr\o{}der}}, \bibinfo {author} {\bibfnamefont {U.~R.}\ \bibnamefont
  {Pedersen}}, \bibinfo {author} {\bibfnamefont {N.~P.}\ \bibnamefont
  {Bailey}},\ and\ \bibinfo {author} {\bibfnamefont {J.~C.}\ \bibnamefont
  {Dyre}},\ }\bibfield  {title} {\bibinfo {title} {Pressure-energy correlations
  in liquids. iv. “isomorphs” in liquid phase diagrams},\ }\href
  {https://doi.org/http://dx.doi.org/10.1063/1.3265957} {\bibfield  {journal}
  {\bibinfo  {journal} {J. Chem. Phys.}\ }\textbf {\bibinfo {volume} {131}},\
  \bibinfo {eid} {234504} (\bibinfo {year} {2009})}\BibitemShut {NoStop}%
\bibitem [{\citenamefont {B\o{}hling}\ \emph {et~al.}(2012)\citenamefont
  {B\o{}hling}, \citenamefont {Ingebrigtsen}, \citenamefont {Grzybowski},
  \citenamefont {Paluch}, \citenamefont {Dyre},\ and\ \citenamefont
  {Schr\o{}der}}]{Boehling2012}%
  \BibitemOpen
  \bibfield  {author} {\bibinfo {author} {\bibfnamefont {L.}~\bibnamefont
  {B\o{}hling}}, \bibinfo {author} {\bibfnamefont {T.~S.}\ \bibnamefont
  {Ingebrigtsen}}, \bibinfo {author} {\bibfnamefont {A.}~\bibnamefont
  {Grzybowski}}, \bibinfo {author} {\bibfnamefont {M.}~\bibnamefont {Paluch}},
  \bibinfo {author} {\bibfnamefont {J.~C.}\ \bibnamefont {Dyre}},\ and\
  \bibinfo {author} {\bibfnamefont {T.~B.}\ \bibnamefont {Schr\o{}der}},\
  }\bibfield  {title} {\bibinfo {title} {Scaling of viscous dynamics in simple
  liquids: theory, simulation and experiment},\ }\href
  {http://stacks.iop.org/1367-2630/14/i=11/a=113035} {\bibfield  {journal}
  {\bibinfo  {journal} {New J. Phys.}\ }\textbf {\bibinfo {volume} {14}},\
  \bibinfo {pages} {113035} (\bibinfo {year} {2012})}\BibitemShut {NoStop}%
\bibitem [{\citenamefont {Schr\o{}der}\ and\ \citenamefont
  {Dyre}(2014)}]{SchroederDyre14}%
  \BibitemOpen
  \bibfield  {author} {\bibinfo {author} {\bibfnamefont {T.~B.}\ \bibnamefont
  {Schr\o{}der}}\ and\ \bibinfo {author} {\bibfnamefont {J.~C.}\ \bibnamefont
  {Dyre}},\ }\bibfield  {title} {\bibinfo {title} {Simplicity of condensed
  matter at its core: Generic definition of a roskilde-simple system},\ }\href
  {https://doi.org/10.1063/1.4901215} {\bibfield  {journal} {\bibinfo
  {journal} {J. Chem. Phys.}\ }\textbf {\bibinfo {volume} {141}},\ \bibinfo
  {pages} {204502} (\bibinfo {year} {2014})}\BibitemShut {NoStop}%
\bibitem [{\citenamefont {Kob}\ and\ \citenamefont {Andersen}(1994)}]{kob94}%
  \BibitemOpen
  \bibfield  {author} {\bibinfo {author} {\bibfnamefont {W.}~\bibnamefont
  {Kob}}\ and\ \bibinfo {author} {\bibfnamefont {H.~C.}\ \bibnamefont
  {Andersen}},\ }\bibfield  {title} {\bibinfo {title} {Scaling behavior in the
  $\ensuremath{\beta}$-relaxation regime of a supercooled lennard-jones
  mixture},\ }\href {https://doi.org/10.1103/PhysRevLett.73.1376} {\bibfield
  {journal} {\bibinfo  {journal} {Phys. Rev. Lett.}\ }\textbf {\bibinfo
  {volume} {73}},\ \bibinfo {pages} {1376} (\bibinfo {year}
  {1994})}\BibitemShut {NoStop}%
\bibitem [{\citenamefont {Bailey}\ \emph {et~al.}(2017)\citenamefont {Bailey},
  \citenamefont {Ingebrigtsen}, \citenamefont {Hansen}, \citenamefont
  {Veldhorst}, \citenamefont {Bøhling}, \citenamefont {Lemarchand},
  \citenamefont {Olsen}, \citenamefont {Bacher}, \citenamefont {Costigliola},
  \citenamefont {Pedersen}, \citenamefont {Larsen}, \citenamefont {Dyre},\ and\
  \citenamefont {Schrøder}}]{RUMD}%
  \BibitemOpen
  \bibfield  {author} {\bibinfo {author} {\bibfnamefont {N.~P.}\ \bibnamefont
  {Bailey}}, \bibinfo {author} {\bibfnamefont {T.~S.}\ \bibnamefont
  {Ingebrigtsen}}, \bibinfo {author} {\bibfnamefont {J.~S.}\ \bibnamefont
  {Hansen}}, \bibinfo {author} {\bibfnamefont {A.~A.}\ \bibnamefont
  {Veldhorst}}, \bibinfo {author} {\bibfnamefont {L.}~\bibnamefont {Bøhling}},
  \bibinfo {author} {\bibfnamefont {C.~A.}\ \bibnamefont {Lemarchand}},
  \bibinfo {author} {\bibfnamefont {A.~E.}\ \bibnamefont {Olsen}}, \bibinfo
  {author} {\bibfnamefont {A.~K.}\ \bibnamefont {Bacher}}, \bibinfo {author}
  {\bibfnamefont {L.}~\bibnamefont {Costigliola}}, \bibinfo {author}
  {\bibfnamefont {U.~R.}\ \bibnamefont {Pedersen}}, \bibinfo {author}
  {\bibfnamefont {H.}~\bibnamefont {Larsen}}, \bibinfo {author} {\bibfnamefont
  {J.~C.}\ \bibnamefont {Dyre}},\ and\ \bibinfo {author} {\bibfnamefont
  {T.~B.}\ \bibnamefont {Schrøder}},\ }\bibfield  {title} {\bibinfo {title}
  {{RUMD: A general purpose molecular dynamics package optimized to utilize GPU
  hardware down to a few thousand particles}},\ }\href
  {https://doi.org/10.21468/SciPostPhys.3.6.038} {\bibfield  {journal}
  {\bibinfo  {journal} {SciPost Phys.}\ }\textbf {\bibinfo {volume} {3}},\
  \bibinfo {pages} {038} (\bibinfo {year} {2017})}\BibitemShut {NoStop}%
\bibitem [{\citenamefont {Ransom}\ \emph {et~al.}(2019)\citenamefont {Ransom},
  \citenamefont {Casalini}, \citenamefont {Fragiadakis},\ and\ \citenamefont
  {Roland}}]{Ransom2019}%
  \BibitemOpen
  \bibfield  {author} {\bibinfo {author} {\bibfnamefont {T.~C.}\ \bibnamefont
  {Ransom}}, \bibinfo {author} {\bibfnamefont {R.}~\bibnamefont {Casalini}},
  \bibinfo {author} {\bibfnamefont {D.}~\bibnamefont {Fragiadakis}},\ and\
  \bibinfo {author} {\bibfnamefont {C.~M.}\ \bibnamefont {Roland}},\ }\bibfield
   {title} {\bibinfo {title} {The complex behavior of the “simplest”
  liquid: Breakdown of density scaling in tetramethyl tetraphenyl
  trisiloxane},\ }\href {https://doi.org/10.1063/1.5121021} {\bibfield
  {journal} {\bibinfo  {journal} {J. Chem. Phys.}\ }\textbf {\bibinfo {volume}
  {151}},\ \bibinfo {pages} {174501} (\bibinfo {year} {2019})}\BibitemShut
  {NoStop}%
\bibitem [{Sup()}]{Supp}%
  \BibitemOpen
  \href@noop {} {\bibinfo {title} {See supplemental material at [url will be
  inserted by publisher]}}\BibitemShut {NoStop}%
\bibitem [{\citenamefont {Ingebrigtsen}\ \emph {et~al.}(2012)\citenamefont
  {Ingebrigtsen}, \citenamefont {Bøhling}, \citenamefont {Schrøder},\ and\
  \citenamefont {Dyre}}]{Ingebrigtsen2012b}%
  \BibitemOpen
  \bibfield  {author} {\bibinfo {author} {\bibfnamefont {T.~S.}\ \bibnamefont
  {Ingebrigtsen}}, \bibinfo {author} {\bibfnamefont {L.}~\bibnamefont
  {Bøhling}}, \bibinfo {author} {\bibfnamefont {T.~B.}\ \bibnamefont
  {Schrøder}},\ and\ \bibinfo {author} {\bibfnamefont {J.~C.}\ \bibnamefont
  {Dyre}},\ }\bibfield  {title} {\bibinfo {title} {Communication:
  Thermodynamics of condensed matter with strong pressure-energy
  correlations},\ }\href {https://doi.org/10.1063/1.3685804} {\bibfield
  {journal} {\bibinfo  {journal} {J. Chem. Phys.}\ }\textbf {\bibinfo {volume}
  {136}},\ \bibinfo {pages} {061102} (\bibinfo {year} {2012})}\BibitemShut
  {NoStop}%
\bibitem [{\citenamefont {Costigliola}\ \emph {et~al.}(2018)\citenamefont
  {Costigliola}, \citenamefont {Pedersen}, \citenamefont {Heyes}, \citenamefont
  {Schrøder},\ and\ \citenamefont {Dyre}}]{Costigliola2018}%
  \BibitemOpen
  \bibfield  {author} {\bibinfo {author} {\bibfnamefont {L.}~\bibnamefont
  {Costigliola}}, \bibinfo {author} {\bibfnamefont {U.~R.}\ \bibnamefont
  {Pedersen}}, \bibinfo {author} {\bibfnamefont {D.~M.}\ \bibnamefont {Heyes}},
  \bibinfo {author} {\bibfnamefont {T.~B.}\ \bibnamefont {Schrøder}},\ and\
  \bibinfo {author} {\bibfnamefont {J.~C.}\ \bibnamefont {Dyre}},\ }\bibfield
  {title} {\bibinfo {title} {Communication: Simple liquids’ high-density
  viscosity},\ }\href {https://doi.org/10.1063/1.5022058} {\bibfield  {journal}
  {\bibinfo  {journal} {J. Chem. Phys.}\ }\textbf {\bibinfo {volume} {148}},\
  \bibinfo {pages} {081101} (\bibinfo {year} {2018})}\BibitemShut {NoStop}%
\bibitem [{\citenamefont {Costigliola}\ \emph {et~al.}(2019)\citenamefont
  {Costigliola}, \citenamefont {Heyes}, \citenamefont {Schrøder},\ and\
  \citenamefont {Dyre}}]{Costigliola2019}%
  \BibitemOpen
  \bibfield  {author} {\bibinfo {author} {\bibfnamefont {L.}~\bibnamefont
  {Costigliola}}, \bibinfo {author} {\bibfnamefont {D.~M.}\ \bibnamefont
  {Heyes}}, \bibinfo {author} {\bibfnamefont {T.~B.}\ \bibnamefont
  {Schrøder}},\ and\ \bibinfo {author} {\bibfnamefont {J.~C.}\ \bibnamefont
  {Dyre}},\ }\bibfield  {title} {\bibinfo {title} {Revisiting the
  stokes-einstein relation without a hydrodynamic diameter},\ }\href
  {https://doi.org/10.1063/1.5080662} {\bibfield  {journal} {\bibinfo
  {journal} {J. Chem. Phys.}\ }\textbf {\bibinfo {volume} {150}},\ \bibinfo
  {pages} {021101} (\bibinfo {year} {2019})}\BibitemShut {NoStop}%
\bibitem [{\citenamefont {Rahman}\ \emph {et~al.}(2021)\citenamefont {Rahman},
  \citenamefont {Carter}, \citenamefont {Saw}, \citenamefont {Douglass},
  \citenamefont {Costigliola}, \citenamefont {Ingebrigtsen}, \citenamefont
  {Schrøder}, \citenamefont {Pedersen},\ and\ \citenamefont
  {Dyre}}]{Rahman2021}%
  \BibitemOpen
  \bibfield  {author} {\bibinfo {author} {\bibfnamefont {M.}~\bibnamefont
  {Rahman}}, \bibinfo {author} {\bibfnamefont {B.~M. G.~D.}\ \bibnamefont
  {Carter}}, \bibinfo {author} {\bibfnamefont {S.}~\bibnamefont {Saw}},
  \bibinfo {author} {\bibfnamefont {I.~M.}\ \bibnamefont {Douglass}}, \bibinfo
  {author} {\bibfnamefont {L.}~\bibnamefont {Costigliola}}, \bibinfo {author}
  {\bibfnamefont {T.~S.}\ \bibnamefont {Ingebrigtsen}}, \bibinfo {author}
  {\bibfnamefont {T.~B.}\ \bibnamefont {Schrøder}}, \bibinfo {author}
  {\bibfnamefont {U.~R.}\ \bibnamefont {Pedersen}},\ and\ \bibinfo {author}
  {\bibfnamefont {J.~C.}\ \bibnamefont {Dyre}},\ }\bibfield  {title} {\bibinfo
  {title} {Isomorph invariance of higher-order structural measures in four
  lennard–jones systems},\ }\href {https://doi.org/10.3390/molecules26061746}
  {\bibfield  {journal} {\bibinfo  {journal} {Molecules}\ }\textbf {\bibinfo
  {volume} {26}},\ \bibinfo {pages} {1746} (\bibinfo {year}
  {2021})}\BibitemShut {NoStop}%
\end{thebibliography}%

\end{document}